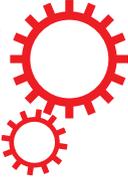

# Rupture Forces among Human Blood Platelets at different Degrees of Activation

Thi-Huong Nguyen[1,2], Raghavendra Palankar[1,2], Van-Chien Bui[1,2], Nikolay Medvedev[1], Andreas Greinacher[2] & Mihaela Delcea[1,2]

Little is known about mechanics underlying the interaction among platelets during activation and aggregation. Although the strength of a blood thrombus has likely major biological importance, no previous study has measured directly the adhesion forces of single platelet-platelet interaction at different activation states. Here, we filled this void first, by minimizing surface mediated platelet-activation and second, by generating a strong adhesion force between a single platelet and an AFM cantilever, preventing early platelet detachment. We applied our setup to measure rupture forces between two platelets using different platelet activation states, and blockade of platelet receptors. The rupture force was found to increase proportionally to the degree of platelet activation, but reduced with blockade of specific platelet receptors. Quantification of single platelet-platelet interaction provides major perspectives for testing and improving biocompatibility of new materials; quantifying the effect of drugs on platelet function; and assessing the mechanical characteristics of acquired/inherited platelet defects.

Blood platelets are discoidal, anuclear cell fragments, $1-2\mu m$ in size and are produced from the cytoplasm of bone marrow megakaryocytes[1]. They prevent blood loss upon vascular injury[2], can induce arterial thrombosis[3], and have additional roles in immune defense[4], wound healing[5], and cancer metastasis[6,7]. *In vivo*, platelets do not normally interact with the inner surfaces of blood vessels, but they adhere promptly when the vessel wall is altered (*e.g.* GPIbIX with von Willebrand factor [vWF], GPIaIIa and GPVI with collagen) leading to platelet activation[1]. Activated platelets expose GPIIbIIIa ($\alpha IIb\beta 3$) in an activated form, which allows binding of fibrinogen[8,9] and triggers platelet activation and aggregation[10]. In parallel, activated platelets release prothrombotic substances from their granules[11]. Some of these substances recruit and activate neighboring platelets, while others enhance thrombin generation together with plasma clotting factors. Finally, this leads to formation of a hemostatic plug at the site of endothelial damage which eventually results in blood vessel closure[12,13].

An increasingly important issue for biotechnological applications in medicine is modulation of platelet activation on various surfaces[14,15]. However, these processes are not well understood as platelets are difficult to handle because they activate immediately after short contact with non-physiological surfaces[16–18], at elevated shear stress[19] and upon air contact[20]. Among these, platelet activation on surfaces is not only a major drawback for microfluidic devices[21], micro- and nano- particulate drug delivery systems[22,23], and intravascular devices[24], but it is also essential for developing tools for direct measuring of platelet mechanics[25,26].

Platelet-surface interactions have been quantified in real-time using calcium mobilization assays, showing that the lag time until platelets activate depends on the physico-chemical properties of the surface[27]. It has been shown that platelets are activated nearly instantaneously upon coming in contact with immobilized fibrinogen and completing their contraction after 15 min[28]. Others have provided information regarding morphology[17,18,29], and elastic moduli[29–31] of activated platelets under physiological conditions. Although these studies have successfully identified significant characteristics of platelets on a single molecular level, none of these previous studies describes quantitatively single platelet-platelet interactions, neither in activated, nor in non-/weakly activated states.

[1]Nanostructure Group, ZIK HIKE - Center for Innovation Competence, Humoral Immune Reactions in Cardiovascular Diseases, University of Greifswald, 17489 Greifswald, Germany. [2]Institute for Immunology and Transfusion Medicine, University Medicine Greifswald, 17475 Greifswald, Germany. Correspondence and requests for materials should be addressed to T.-H.N. (email: thihuong.nguyen@uni-greifswald.de) or A.G. (email: greinach@uni-greifswald.de) or M.D. (email: delceam@uni-greifswald.de)

Single-cell force spectroscopy (SCFS) experiments measure adhesive interaction forces and binding kinetics under physiologically relevant conditions at single cell regimes and contribution of single or a few molecules to such interactions[32]. SCFS is increasingly used to study ligand-receptor interactions in living cells[33–35], microbial cell adhesion[36], integrin and glycocalyx mediated contributions to cell adhesion[37], as well as single cell-cell interaction[32,38]. For measurement of interaction forces between two cells with SCFS, a single cell is immobilized at the end of a cantilever and another cell is fixed on a solid substrate. The cell on the cantilever is brought into contact with the cell on the substrate for interaction, and the rupture force is measured when the cells are separated from each other. One of the difficulties in measuring cell-cell interaction is the weak binding strength between cells and cantilever/substrate, leading to an early cell detachment, and therefore, the cantilever/substrate must be modified to increase cell/surface adhesion. For platelets, however, substrate modification with commonly used cell adhesion materials can lead to rapid platelet spreading and activation. To date, almost all studies involving surface modification have been performed with activated platelets. Minimizing non-specific platelet-surface activation is an urgent need not only for platelet mechanics, but also for designing platelet function testing miniaturized microfluidic devices.

In this study we describe using SCFS a strategy to achieve strong adhesion of a single platelet to an atomic force microscopy cantilever (AFM-cantilever) with no-/or minimal platelet activation, which allows measurements of rupture forces among single platelets. We first monitored platelet activation levels by different materials, that is, collagen, fibronectin, and poly-L-lysine, and determined the most suitable material to modify the AFM-cantilever for immobilization of a single platelet. Individual response of platelets to different material-passivated substrates was utilized to measure the rupture forces among single platelets at different activation states. We have successfully applied our approach to two different systems, that is, native and modified platelets. Our approach has potential application for assessing the interaction of platelets with novel material-based implant surfaces, the mechanical characteristics of acquired/inherited platelet defects, the interactions of platelets with other cells, as well as for biotechnology and pharmaceutical developments.

## Results

**Platelet activation induced by different materials.** The degree of platelet activation induced by collagen-, fibronectin-, and poly-L-lysine (PLL)-passivated glass surfaces was characterized by comparing platelet morphologies, P-selectin expression levels, and calcium mobilizations. Two types of collagen were used: i) collagen G type I (here called Collagen G), an acid-soluble calfskin collagen, and ii) equine collagen type I reagens Horm (here called Horm collagen). Collagen G is frequently used to create a gel-like substrate for cell culture, while Horm collagen is used as an agonist to induce platelet activation and aggregation in platelet function tests. For all surfaces, scanning electron microscopy (SEM) showed that platelets activate less after 1 min surface contact, and they reach a higher degree of activation after longer contact times (5-, 10-, 15- and 35 min), as indicated by the loss of their discoidal shape and formation of filopodia (Fig. 1 and Supplementary Fig. 1). At 15 min after surface contact (a sufficient time frame to complete a SCFS measurement), platelets showed extensive spreading and numerous filopodia on PLL (Fig. 1A) and also on Horm collagen (Fig. 1B) indicating a high degree of activation. However, the activation was weaker on fibronectin (Fig. 1C), and weakest on collagen G (Fig. 1D). At longer surface contact time (35 min), we observed some activation of platelets on collagen G, but highly activated platelets on fibronectin, Horm collagen, and PLL (Supplementary Fig. 1). The degree of platelet activation due to platelet-surface interaction was also measured based on P-selectin expression levels by immunofluorescence labelling (Fig. 1E–H). P-selectin expression was lowest on collagen G (grey, Fig. 1I) in comparison with fibronectin (red, Fig. 1I), PLL (blue, Fig. 1I), and Horm collagen (magenta, Fig. 1I). In addition, single platelet spreading areas on different passivated substrates were quantified. The area of platelet spreading was smallest on collagen G compared to platelets spreading on Horm collagen, fibronectin, and PLL (Fig. 1J). The lower spreading level also indicates that platelets are less activated on collagen G than on other passivated substrates, which is consistent with the measurement of P-selectin expression levels and observations of morphology by SEM.

Furthermore, platelet aggregation induced by different materials in solution was tested by platelet aggregometry using both platelet-rich plasma and washed platelets. With platelet-rich plasma, Horm collagen induced immediately platelet aggregation at concentration of $5\,\mu g/ml$, while collagen G did not (Supplementary Fig. 2A). However, collagen G at 20 times higher concentration ($100\,\mu g/ml$) did induce platelet aggregation but still after a much longer lag phase, while PLL did not induce platelet aggregation. With washed platelets (Supplementary Fig. 2B), Horm collagen also induced immediately platelet aggregation at concentration of $5\,\mu g/ml$, while collagen G only induced platelet aggregation at 20 times higher concentration ($100\,\mu g/ml$). With washed platelets, the lag time until platelet aggregation induced by collagen G ($100\,\mu g/ml$) was also shorter compared to the lag time in platelet rich plasma. These results indicate that equine Horm collagen induces platelet aggregation much stronger than collagen G. Whether this is due to different interaction of the different collagens with the platelet receptors GPIa/IIa and GPVI, or different interaction of von Willebrand factor and these collagens[39] requires further studies. PLL ($100\,\mu g/ml$) induced aggregation of washed platelets and reached a similar maximum as thrombin receptor activating peptide (TRAP 6) or Horm collagen but at much shorter lag phase ($34.2\,s$ vs $53.2\,s$ and $68.2\,s$, respectively). PLL induced aggregation in washed platelets but not in platelet-rich plasma, which may be due to the fact that PLL charges were neutralized by proteins available in plasma. These observations are consistent with Guccione et al.[40] who showed that PLL caused aggregation after a longer lag phase in platelet-rich plasma. In addition, we performed calcium mobilization imaging using Fluo-4 AM loaded platelets and observed minimal platelet spreading and delayed release and mobilization of calcium in platelets on collagen G, but maximal mobilization of calcium on PLL (Supplementary Fig. 3 and Supplementary Movies 1–4). Overall, we conclude that platelet activation is weakest on collagen G, intermediate on fibronectin, and strongest on PLL.

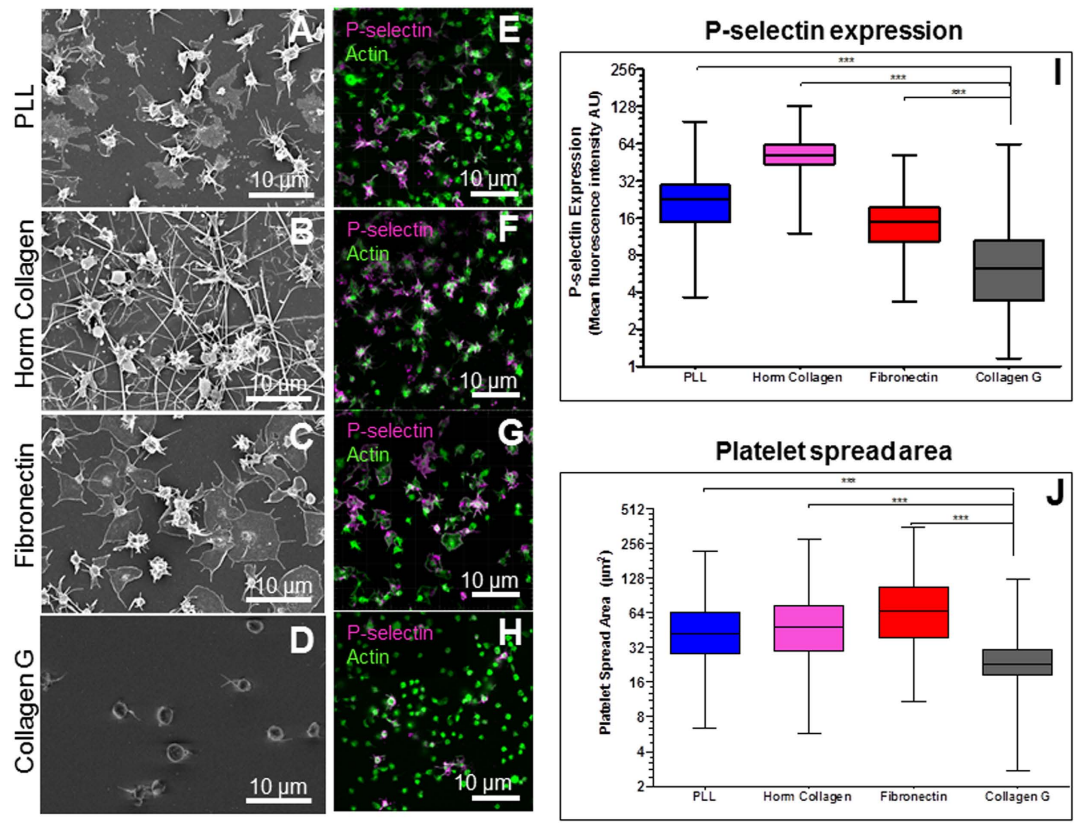

**Figure 1. Characteristics of platelet-surface activation.** SEM togheter with P-selectin and actin indicators show that platelets activate rapidly on PLL (**A,E**) and Horm collagen (**C,G**), slower on fibronectin (**B,F**), while they keep their spherical shape on collagen G (**D,H**) after 15 min surface-contact. (**I**) P-selectin level is lowest for platelets on collagen G (black), higher on fibronectin (red), followed by PLL (blue) and highest on Horm collagen (magenta). (**J**) Spread area of platelets on PLL (blue), fibronectin (red) and Horm collagen (magenta) are significantly higher than on collagen G (black). The thin filopodial extensions and irregular shape are characteristics of activated platelets. Platelets used for calculation: for each experimental condition n = 340 for P-selectin (**I**) and n = 360 for platelet spread area (**J**).

## Establishment of a strong adhesion between platelet and AFM-cantilever.

In order to further understand the interaction behavior of platelets on the investigated materials, we measured the interaction forces between single platelets and different material-passivated substrates. To measure a platelet-probe, a tipless cantilever was first passivated by incubation with collagen G, and then brought into contact with a loosely bound platelet on collagen G passivated substrate for adhesion (Fig. 2 A). After 3–5 min, the cantilever was gently lifted up picking a single platelet from the substrate (Fig. 2B,C). The platelet-probe was then brought into contact (Fig. 2D) with collagen G, fibronectin, and PLL-passivated glass for 15 min to allow the platelet to adhere and spread (Fig. 2E), during which the cantilever is pulled down and bent. This deflection of the cantilever was then converted into spreading force ($F_s$) (Fig. 2E), that was defined as the force required for single platelet adhesion and spreading (or activation) on the material-passivated substrate. After 15 min of contact (Fig. 2F), the spreading force of the platelet on collagen did not significantly change (only $5 \pm 3$ nN) (black trace), while it increased to $23 \pm 5$ nN on fibronectin (red trace), and to $35 \pm 4$ nN on PLL (blue trace). The magnitudes of spreading forces during force field evolution reveal that platelets spread weakest on collagen, stronger on fibronectin, and strongest on PLL. Moreover, the results also indicate that the platelet was still adherent on the collagen-passivated cantilever even though a strong force ($\geq 35 \pm 4$ nN) pulled it away from the cantilever. Since the interaction/rupture force between two cells is typically known of being only several nanonewtons[37], the strong adhesion force between collagen and platelet was sufficient to maintain the platelet on the cantilever during SCFS measurements. We, therefore, selected collagen G as a material to adhere platelets on the AFM-cantilevers for single platelet-platelet interaction measurements.

## Rupture forces between platelets at different states of activation.

Our findings were first applied to measure the rupture force among native platelets. We measured the interactions of three platelet-platelet pairs: i) a non-/weakly activated platelet with a non-/weakly activated platelet, ii) a non-/weakly activated platelet with a partially activated platelet, and iii) a non-/weakly activated platelet with an activated platelet. To generate different degrees of activation, platelets were immobilized on a glass surface passivated either with collagen G, fibronectin, and PLL as substrates to form non-/weakly-, partially-, and strongly activated platelets, respectively. The rupture force between two single platelets was measured by approaching the platelet-probe to another platelet on material-passivated substrate (Fig. 3A–C), followed by the retraction of the cantilever from the substrate. With this setup, the interaction forces between two single platelets, in which the platelets on the passivated substrates exposed different degrees of activation, were measured. As a control, a platelet agonist TRAP 6 was added to the set-up in which platelets were immobilized on both cantilever and substrate *via* collagen G.

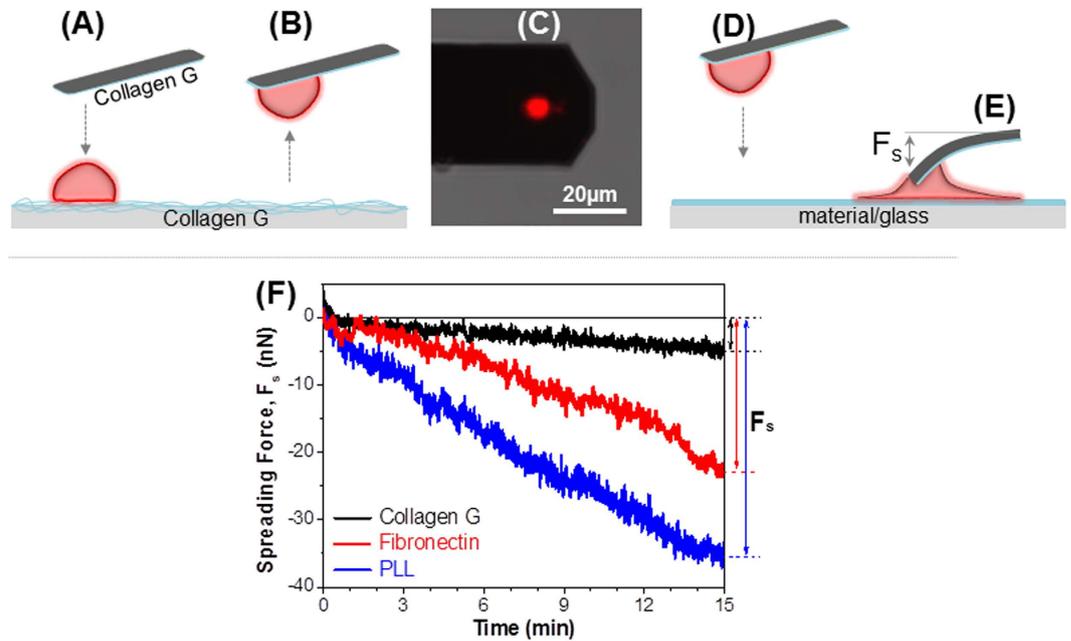

**Figure 2. Single platelet spreading within 15 min on material-passivated substrates. (A)** Collagen G-passivated tipless cantilever is brought to a loosely bound platelet (**red**) on collagen G for adhesion. (**B**) Cantilever with a single platelet is slowly moved up from the substrate. (**C**) Confocal laser scanning microscopy image of the single platelets stained with Vybrant DiD ex. 644/em.665 (red) at the end of an AFM cantilever. (**D**) Platelet-probe is brought into contact with material-passivated glass. (**E**) During activation, platelet spreads and pulls down the cantilever resulting in a spreading force ($F_s$). (**F**) The spreading force induced by platelets is low on collagen G (**black trace**), higher on fibronectin (**red trace**), and highest on PLL (**blue trace**).

Platelet activation leads to release of dense and alpha-granules packed with a wide variety of molecules, collectively called as the platelet secretome consisting up to 300 distinct proteins[11,41]. These proteins (called here as mediators, dark cyan strings between two platelets, Fig. 3D–F) upon release from activated platelets play a key role as bridging molecules between platelets (such as fibrinogen and vWF)[10,42]. The degree of platelet activation influenced by different agonists may determine the type and quantity of the released mediators, which in turn, might result in the differences in the magnitudes of binding forces observed. In addition platelet activation results in inside-out signaling with activation of integrins, which in turn bind more bridging molecules. According to the results of aggregometry, P-selectin expression, calcium signaling, and platelet morphology, we assume that the number of mediators between two platelets is lowest on collagen (Fig. 3D), higher on fibronectin (Fig. 3E), and highest on PLL (Fig. 3F). Figure 3G presents typical retraction curves for platelet-platelet interaction obtained when platelets were immobilized on collagen (black), fibronectin (red), and PLL (blue). The short range adhesion force ($F_{SR}$) shows a maximum interaction between two single platelets, while the saw-tooth like parts ranging from 0 μm to about 6 μm indicate the rupture of mediators between membrane receptors of the platelet-platelet pairs. The different force heights of the saw-tooth like peaks could be attributed to the ruptures of multiple and possibly dissimilar proteins between two platelets. The long range rupture force ($F_{LR}$), *i.e.*, the final saw-tooth like point, is the rupture force of the last mediator between two platelets before they completely separate from each other. Here, we analyzed the short range adhesion force ($F_{SR}$) to quantify the maximal interaction forces between single platelets.

In order to quantify the interaction force between single platelets, we recorded at least 500 force-distance (F-D) curves from 7 to 10 single platelet pairs for each experiment within 15 min on each passivated substrate and collected the adhesion forces ($F_{SR}$) in a histogram for analysis. The average adhesion force values and their corresponding errors were obtained by fitting the histogram with Gaussian fits (Fig. 4A) for platelets on collagen (black), platelets on collagen in the presence of TRAP 6 (grey), fibronectin (red), and PLL (blue). The interaction force of the platelet/platelet pairs (Fig. 4B) was lowest on collagen ($1.50 \pm 0.05$ nN) (black), higher on TRAP 6 ($2.1 \pm 0.33$ nN) (grey), fibronectin ($2.01 \pm 0.05$ nN) (red), and highest on PLL ($2.61 \pm 0.04$ nN) (blue).

However, both long and short range forces do not describe the whole interaction landscape of all mediators between two platelets. After rupturing at the maximal force ($F_{SR}$), rupture of mediators still occurred during platelet separation, as indicated by the saw-tooth like structures (Fig. 3G). These ruptures occurred in between the short and long range interactions. In order to better quantify all mediators which appeared between two platelets, we analyzed the dissipation/interaction work (W) using the area under the curve below the zero line (shaded area, W, Fig. 5A). The energy dissipated during this lift-off process was the shaded area confined by the starting point (distance = 0, force = 0) and the final detachment event (distance = D, force = F). The average interaction work was calculated by integrating interaction force (f) at each individual rupture distance (d): $W = \int_0^D f \times d^{43,44}$. The calculation was provided by the JPK data processing software.

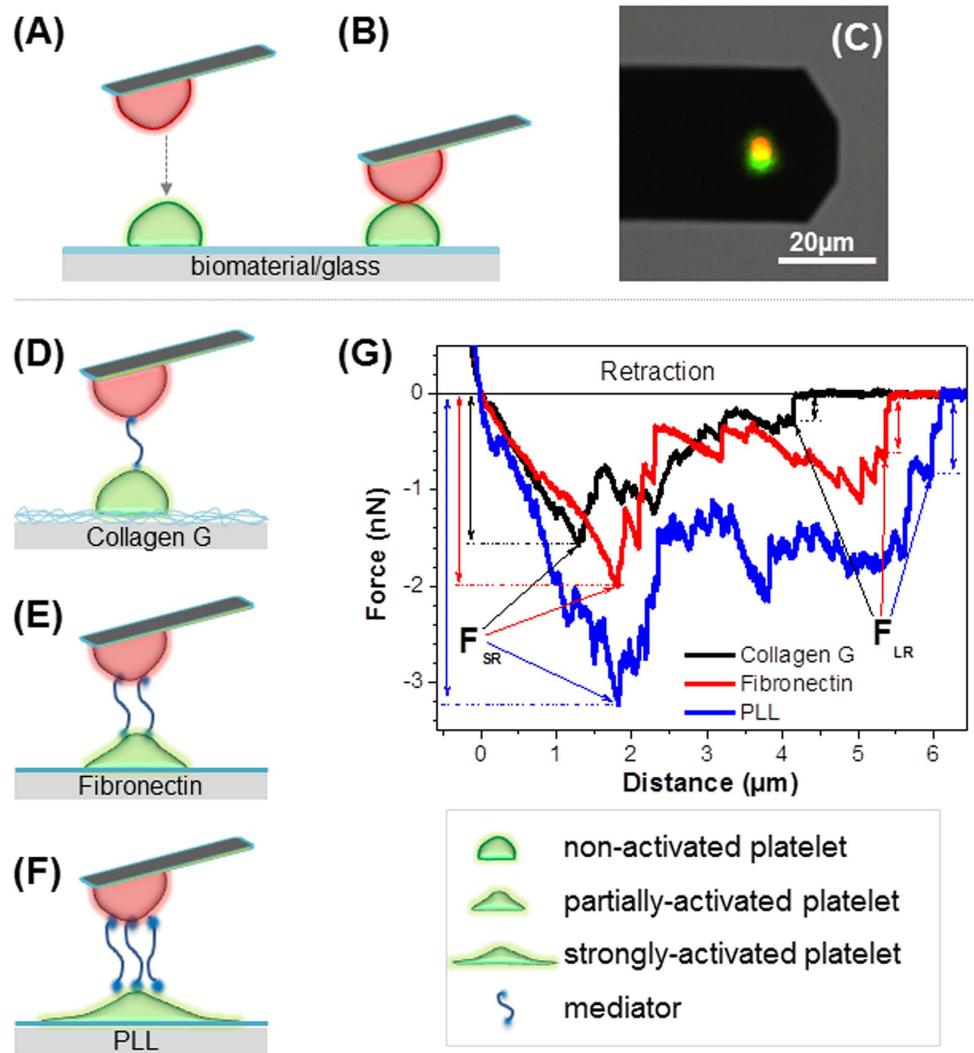

**Figure 3. Schematic illustration of single cell force spectroscopy.** A single platelet (**red, a**) is brought into contact with another platelet on the substrate (**green, A**) for interaction (**B**) as indicated by merging of two colors in the confocal laser scanning microscopy image (**C**). Cartoon of platelet-platelet interaction, when the platelet on the substrate is adhered to different materials, which results in different degrees of platelet activation and/or release of different molecules (named mediators) (**dark cyan in D-F**). Mediators become adhesive factors between the platelet on the cantilever and the platelet on the substrate: lowest level on collagen G (**D**), higher on fibronectin (**E**), and highest on PLL (**F**). When the cantilever is separated from the substrate, the adhesion force between two platelets will be measured. (**G**) Typical retraction curves obtained when the platelet on the substrate is attached to collagen G (**black**), fibronectin (**red**), or PLL (**blue**) showing the short range adhesion forces ($F_{SR}$) and long range adhesion forces ($F_{LR}$).

The dissipation work of platelet-platelet pairs (Fig. 5B) was found to be lowest on collagen G ($1.40 \times 10^{-16} \pm 0.04 \times 10^{-16}$ J) (black), higher in the presence of TRAP 6 ($2.72 \times 10^{-16} \pm 0.07 \times 10^{-16}$ J) (grey), on fibronectin ($2.12 \times 10^{-16} \pm 0.05 \times 10^{-16}$ J) (red), and highest on PLL ($3.73 \times 10^{-16} \pm 0.08 \times 10^{-16}$ J) (blue). The dissipation work during rupture of platelet-platelet pairs followed the same pattern as observed for the interaction force (described in Fig. 4), that is, lowest on collagen, higher on fibronectin, and highest on PLL. In addition, we observed multiple peaks in the work distribution histogram, which were not possible to recognize from the force distribution histogram for platelets on PLL (Supplementary Fig. 4). These multiple peaks point toward the rupture of several mediators between two platelets.

The different rupture forces or dissipation work were dependent on the level of platelet activation, which proved that our experimental setup allows measurement of differences among single platelets exposing different degrees of activation.

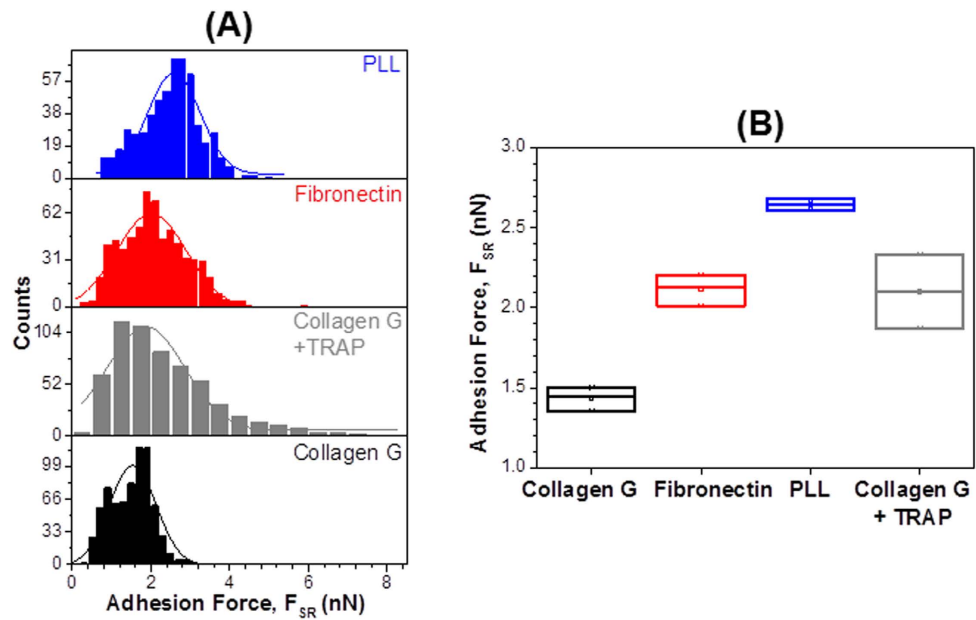

**Figure 4. Rupture forces among single platelets exposing different degrees of activation.** (**A**) Histograms showing rupture forces between the platelet-probe and platelet immobilized on collagen G (black), collagen G in the presence of TRAP 6 (grey), fibronectin (red), and PLL (blue). Solid lines are the corresponding Gaussian fits to the data. (**B**) Average rupture forces with the corresponding standard errors collected from three independent experiments. $F_{SR}$ = short range forces.

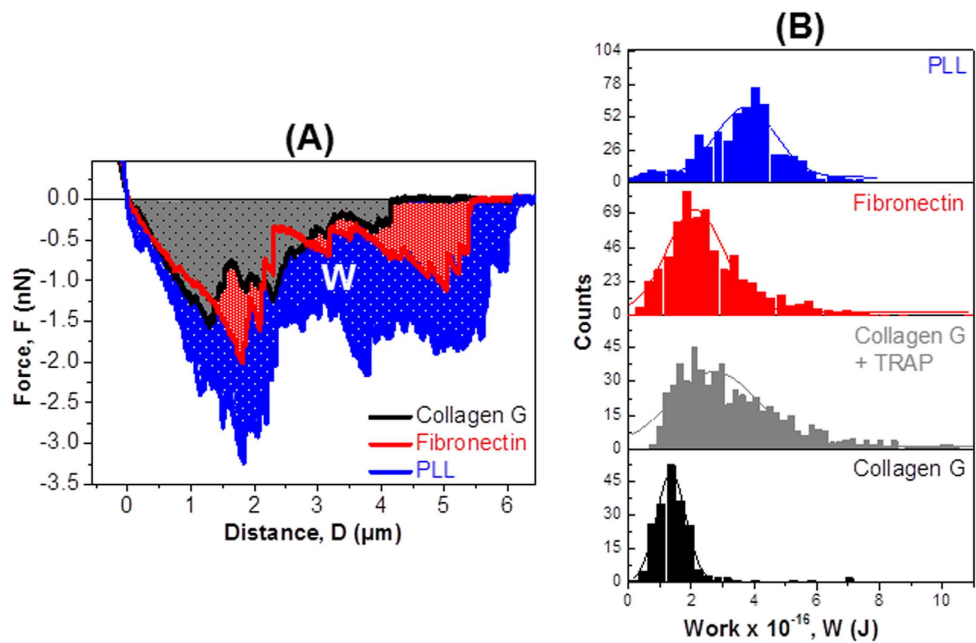

**Figure 5. Dissipation works among single platelets exposing different degrees of activation.** (**A**) Dissipation work required to separate the platelet on the cantilever from the platelets on collagen G (black), collagen G in the presence of TRAP 6 (grey), fibronectin (red), or PLL (blue). (**B**) Dissipation work obtained from one typical experiment when the basal platelet is immobilized on collagen G (black), collagen G with TRAP 6 (grey), fibronectin (red), or PLL (blue). Solid lines are Gaussian fits.

**Rupture forces among modified platelets.** We then applied our approach to measure the rupture force among platelets treated with receptor blockers. GPIIbIIIa is the major receptor for fibrinogen and plays the most important role in platelet-platelet interactions, platelet-plug stability in hemostasis and platelet aggregation. We tested whether blocking of GPIIbIIIa receptors using abciximab, a $F_{ab}$ fragment of a monoclonal antibody, which binds to the ligand binding site of GPIIbIIIa, allows the measurement of additional platelet-platelet rupture forces.

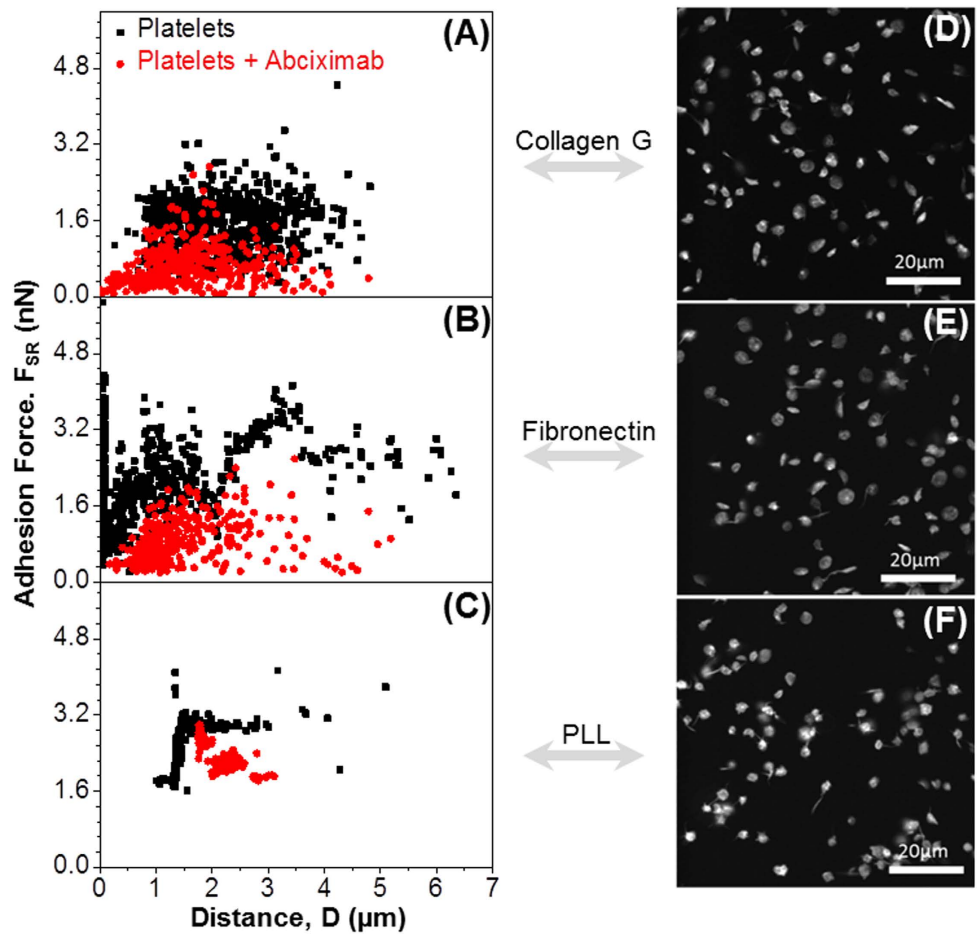

**Figure 6. Rupture forces among single modified platelets.** Rupture force *vs* rupture distance measured in the absence (**black**) and presence (**red**) of abciximab, an inhibitor of GPIIbIIIa. When platelets are preincubated with abciximab, the rupture force reduces on collagen G (**A**), fibronectin (**B**), and PLL (**C**). The corresponding fluorescent images show platelets (cytoskeleton labelled with Phalloidin Atto565) incubated with abciximab on collagen G (**D**), fibronectin (**E**), or PLL (**F**).

We found that the rupture forces between two platelets (red dots, Fig. 6) after blocking GPIIbIIIa receptors, were significantly reduced, when the basal platelet was immobilized on collagen (Fig. 6A) and on fibronectin (Fig. 6B), but less reduced when it was immobilized on PLL (Fig. 6C) as compared to native platelets (black dots, Fig. 6). Besides the reduction of rupture forces using inhibitors, we also observed by confocal microscopy a reduction of platelet spreading on collagen (Fig. 6D), fibronectin (Fig. 6E) and PLL (Fig. 6F) when GPIIbIIIa was blocked.

## Discussion

The present study provides a method to measure forces generated during platelet spreading on different substrates and the rupture forces between two single platelets. Our data provide insight into the interaction among platelets at different degrees of activation. To achieve this, we first had to create a strong adhesion force between single platelets and the AFM-cantilevers to avoid early detachment of the platelet from the cantilever, while in parallel, platelet activation by the cantilever surface had to be minimized.

For SCFS, a time frame from milliseconds to tens of minutes[45] is required to complete a measurement set (F-D curves of ~500 for each individual condition). During this time, platelets may become activated resulting in a change of measured forces. Previous studies have shown that activated platelets on a substrate exhibit a progression of morphological changes from dendritic to fully spread forms[46–48].

Among the investigated substrates (e.g. collagens, fibronectin and PLL), we observed that platelets spread and activated slowest on collagen G, that is, they kept their non-/weakly activated form for up to ~30 min upon surface-contact. It was surprising to find that collagen G-passivated glass substrate slowed down platelet-surface activation, while it is known that collagen is a platelet agonist which interacts with GPIa/IIa (integrin α2β1) and GPVI on the platelet membrane[49]. We therefore, carefully compared the reactivity of platelets with collagen G and with the collagen widely used for platelet activation studies (Horm collagen) by different techniques (aggregometry, P-selectin expression, calcium signaling, and platelet morphology) and found that Horm collagen-passivated substrate strongly activate platelets. The delay in the onset of platelet activation on collagen G when used as a passivating material on substrates becomes an interesting approach for many other applications such as investigation of platelet mechanics in diseases, or microfluidic platelet function testing devices.

We found that the stiffness and thickness of the passivated matrix layers formed by the investigated materials are correlated with spreading of platelets. Indentation depth of the gel formed by collagen G was ~200 nm and only ~100 nm for Horm collagen, while the Young's moduli were $1.561 \pm 0.013$ kPa and $2.315 \pm 0.122$ kPa for collagen G and Horm collagen, respectively (Supplementary Fig. 5). Recently, Kee *et al.* reported that stiffness of collagen-conjugated polyacrylamide gels affects platelet adhesion, spreading, and activation, and showed that the boundary between weak and strong platelet activation is at the stiffness of 5 kPa[50]. Here we also found that platelets spread slowest on the softest surface (collagen G) and faster on stiffer surfaces (Horm collagen, fibronectin and PLL).

Strong platelet spreading and activation on PLL is attributed to the combinational effects of surface stiffness and PLL's functionality in platelet-induced aggregation. Consistent with previous studies[40,51,52], PLL which is frequently used to immobilize cells for studies of cell mechanics, strongly activated platelets due to the electrostatic interaction between platelet membrane and positively charged PLL[51,53]. For a thin coating layer, platelets 'sense' the stiffness of the substrate, and therefore, spread and activate faster on fibronectin or PLL than on collagen G. Potentially, instead of spreading on collagen G gel, platelets penetrate because they are stiffer (1–50 kPa)[30] than collagen G gel (~1 kPa) (Supplementary Fig. 5).

Surprisingly, platelets activate slowly although they bind strongly to collagen G (~35 nN, Fig. 2). The strong binding is consistent with the investigations of Lam *et al.*[28]. Most probably platelets settle on collagen networks and form pseudopodia at the bottom of the platelet, which "anchor" in the corresponding tiny holes of the collagen G network (Supplementary Fig. 5B). The high force magnitude (~35 nN) is sufficient to maintain the platelet on the cantilever during SCFS measurements since the interaction forces between two cells were normally about only several nanonewtons. At the same time, platelets did not show typical signs of major shape change, spreading and activation for a period of about 30 min. We, therefore, selected collagen to adhere platelets on the AFM-cantilevers for single platelet-platelet interaction measurements.

During thrombus formation, not only the first layer of platelets binding to a thrombotic surface matters, but also the subsequent interactions with surrounding platelets and other blood cells (e.g., leukocytes) which play a role in hemostasis and innate immune defense mechanisms. Our method allows addressing these secondary interactions by measuring the rupture force between two single platelets. We first applied this approach to quantify the rupture forces of the native platelet-platelet pairs, in which a single platelet was always attached to the cantilever and the other platelet was immobilized on substrates passivated with different materials. We found the lowest rupture force when the platelet was immobilized on collagen-passivated substrate, higher on fibronectin, and highest on PLL. These different rupture forces are a composite of the activation state of platelet surface receptors and binding partners of these receptors, such as von Willebrand factor, vitronectin, or fibrinogen. Since the SCFS experiments were performed with washed platelets, there were no additional plasma proteins and therefore, proteins involved had to be released during platelet activation, e.g. from the platelet alpha-granules[10]. It is therefore, no surprise that the magnitude of the platelet-platelet interaction forces correlated with the level of platelet activation.

To show the potential of SCFS to unravel single parts of the multiple and complex platelet-platelet interactions, we next specifically inhibited one type of platelet receptor, i.e., GPIIbIIIa which is the major integrin on the platelet surface (~80 000 copies/platelet)[54] and plays the most important role in platelet aggregation. We blocked the GPIIbIIIa receptors using abciximab (a $F_{ab}$ fragment of a monoclonal antibody, which blocks the ligand binding site of the GPIIbIIIa). Not surprisingly, the rupture forces between two platelets after blocking GPIIbIIIa receptors were strongly reduced. However, interaction forces were still present. The results show that specific blocking of targeted receptors on the platelet surface allows measuring of the interactions of platelets *via* other receptors than GPIIbIIIa, and further underscore the power of SCFS to characterize the interactions between single platelets and platelets with artificial surfaces, as well as its potential to quantify the impact of pharmacologic drugs on platelet mechanics.

We have described the change in platelet-platelet interaction in terms of adhesion force. However, analysis of adhesion force does not reflect the whole interaction landscape of all mediators between two platelets. By analyzing the dissipation work (W), we obtained some additional details. First, by considering the interaction between platelets on collagen as a background, we found that the ratios of dissipation work ($R_W = W_{fibronectin} : W_{collagen\ G} \sim 1.5$ and $R'_W = W_{PLL} : W_{collagen\ G} \sim 2.7$) increased considerably more than those of rupture force ($R_F = F_{fibronectin} : F_{collagen\ G} \sim 1.3$ and $R'_F = F_{PLL} : F_{collagen\ G} \sim 1.9$). The increase of the ratios indicates that the impact of mediators between platelets can be better quantified by considering dissipation work as compared to adhesion force. Second, we observed multiple peaks in the work distribution histogram, which were not possible to recognize from the force distribution histogram. These multiple peaks show the rupture of many mediators between two platelets. As outlined above, inhibition of certain receptors with specific reagents, or the use of knock-out platelet models can be applied to further characterize these interactions.

Despite being the first method allowing characterization of the mechanics of platelet-platelet interactions, our approach has some limitations in addition to general limitations occurring in single cell force spectroscopy measurements such as low throughput, time constraints involved and thermal drift during AFM[45]. Moreover, due to gradual activation of platelets ($\geq$30 min), long term kinetic measurements are difficult to interpret mechanistically; hence we limited our SCFS measurement window to $\leq$15 min. We also use the term non-/weakly activated platelets throughout the manuscript because it is likely impossible to prepare washed, immobilized platelets without inducing platelet activation. As assessed by morphologic characteristics and the more sensitive calcium influx, we observed no indication for major platelet activation. Nevertheless, experiments using the described method should be completed within 30 min. Thereafter, artifacts caused by platelet activation/aggregation and molecules released from platelet secretome need to be considered. Second, transferring the platelet-probe to other samples may result in detachment of the platelet from the cantilever due to capillary force (up to hundreds of nN). We are currently working on overcoming this limitation by generating patterned surfaces which allow functionalization of the same substrate with various materials without the need of changing sample.

Our SCFS approach opens a wide range of applications to assess the impact of platelets in biotechnology. It allows measurement of platelet activation by different materials used for intravascular devices (*e.g.* vascular grafts, intravascular stents); the impact of drugs on platelet-surface interaction as well as platelet-platelet interaction. In combination with specific inhibitors as well as targeted knock-out mice for specific platelet proteins, our method enables a wide range of experimental approaches to better understand the interplay between platelets, surfaces, and drugs.

## Methods

**Scanning electron microscopy (SEM) imaging of platelets.**    24 mm round glass coverslips or silicon chips (Plano GmbH, Wetzlar, Germany) were cleaned as previously described[55]. The freshly cleaned substrates were then incubated for 3 h at 37 °C in 50 μg/mL fibronectin or collagen G type I (an acid-soluble calfskin collagens, Biochrom GmbH, Berlin, Germany) or Horm collagen (a collagen Reagens Horm from horse, Takeda, Singen, Germany) or 100 μg/mL Poly-L-Lysine (PLL) molecular weight 70–150 kDa (Sigma Aldrich, Darmstadt, Germany). To investigate the influence of surface roughness induced by collagen G, 50 μg/mL collagen G was incubated at 37 °C overnight. Subsequently, these material-passivated substrates were rinsed three times with phosphate buffered saline (PBS), and dried under nitrogen stream. An aliquot of 10 μl of healthy human donors at a density of $3 \times 10^5$ platelets/μl, washed as previously described[56], were dropped on the freshly-passivated substrates and kept at room temperature (RT) for 1-, 5-, 15-, and 35 min. Fixing and gold-coating of platelets were performed as described[55].

**Calcium mobilization assay and immunofluorescence microscopy.**    Glass-bottom 35 mm dishes (IBIDI, Martinsried, Germany) were exposed to UV cleaner (Bioforce Labs, Ames, IA, USA) for 30 min. Then, they were passivated with collagens (50 μg/ml), or fibronectin (50 μg/ml), or PLL (100 μg/ml) as described above. Subsequently, washed human platelets obtained from healthy volunteers after informed consent ($3 \times 10^5$ platelets/μl) were incubated with 5 μM $Ca^{2+}$ indicator Fluo-4 AM (Life Technologies GmbH, Darmstadt, Germany) for 20 min in dark at RT and washed twice to remove excess dye. Subsequently, 25 μl samples of platelet suspensions of the Fluo-4 AM platelets were pipetted onto the passivated substrates and images were recorded continuously for 15 min with an interval of 6 seconds in sequential scanning mode at RT on a Leica TCS SP5 confocal microscope (Leica GmbH, Wetzlar Germany). For imaging of platelet-platelet pair on the cantilever and on the substrate, platelet on the cantilever was incubated with Vybrant DiD ex. 644/em.665 (red), while platelet on the substrate was incubated with Vybrant DiO ex.488/em.501 (green) (Cat. No.: V-22889, Eugene, OR 97402, United States).

Platelet activation was measured by immunofluorescence detection of P-selectin. Briefly, 200 μL platelet suspension ($15 \times 10^3$/μL) was incubated with passivated glass surfaces for 15 min followed by fixing in 3.7% paraformaldehyde for 30 min and quenching for 5 min in 30 mM glycine/PBS, pH 7.5. Fixed, non-permeabilized platelets were incubated with mouse anti-human CD62P monoclonal antibody (Cat.No. 555522, BD Pharmingen, San Jose, CA, U.S.A) at 20 ng/mL in 0.1% v/v BSA in PBS at 4 °C for 16h followed by incubation with goat anti-mouse Alexa Fluor 647 antibody at 2 μg/mL in 0.1% v/v BSA in PBS at 25 °C for 90 minutes. Phalloidin Atto565 (AD 565–81, ATTO-TEC GmbH, Siegen, Germany) was used at 5 pM in PBS to label platelet actin cytoskeleton. Fluorescence images were acquired on a Leica TCS SP5 confocal laser scanning microscope (Leica GmbH,Wetzlar,Germany). Image processing and quantitation was performed with Imaris (version 7.6.5, Bitplane AG, Zurich, Switzerland), DIP image MATLAB tool box and ImageJ (Rasband, W. S., ImageJ, U.S. National Institutes of Health, Bethesda, MD, 1997_2012, http://imagej.nih.gov/ij). Statistical analysis was carried out on Prism 5 (version 5.03, GraphPad Inc. La Jolla, CA, United States). Statistical comparisons of experimental groups were evaluated by paired 2-tailed Student t test. A $P$ value < 0.05 was considered statistically significant.

**Platelet aggregation assay (Aggregometry).**    Platelet aggregation was monitored by recording changes in light transmission with the use of an aggregometer APACT 4S Plus (Dyasys Greiner Gmb, Flacht, Germany) as previously described[57]. Either platelet-rich plasma prepared as described[58] or washed platelets ($3 \times 10^5$ platelets/μl) were used. TRAP 6 (20 μg/ml) (Bachem, Bubendorf, Switzerland) as a control, or Horm collagen (5 μg/ml), or collagen G (5-, and 100 μg/ml), or PLL (100 μg/ml) was added to platelet suspension at 37 °C and platelet aggregation trace (%) was continuously recorded. For comparison, the same blood donors were used for every set of measurement.

**Determination of surface stiffness.**    A 7.27 μm silica beads (Bangs Laboratories, Indiana, USA) were adhered to a silicon CSC12 tipless cantilevers (MicroMasch, Tallin, Estonia). To avoid adhesion of material-passivated substrates to the bead during measurements, beads were exposed to UV cleaner (Bioforce Labs, Ames, IA, USA) for 30 min prior to incubation with 1% PEG-silane 6–9 units (abcr GmbH, Karlsruhe, Germany). The cantilever spring constants were independently measured by a thermal tune procedure as previously described[59]. Then, force-distance (F-D) curves were performed on bare glass and glass surfaces passivated with 50 μg/ml collagen G or Horm collagen, or 50 μg/ml fibronectin, or 100 μg/ml PLL at a speed 1 μm/s and a setpoint of 300 pN. The Young's modulus was determined by the indentation depth (σ) induced between the slope of the material and the slope of the bare glass applying Hertz model[30,60] using a spherical tip shape, which was available in the JPK software. For each material passivated substrate, 300 F-D curves were recorded at 5 random points over the entire surface, and the average value as well as its corresponding standard error was determined from these measurements by Gaussian fits using Origin software (version 8.6, OriginLab Corporation, Northampton, MA, USA).

**Immobilization of platelet on AFM cantilever.**    Silicon CSC12 tipless cantilevers (MicroMasch, Tallin, Estonia) with a nominal spring constant of 0.6 N/m were first exposed to UV cleaner (Bioforce Labs, Ames, IA, USA) for 30 min. Before coating, the cantilever spring constants were independently measured by a thermal tune procedure as previously described[59]. Then, the cantilevers were incubated in 50 µg/ml collagen G for 3 h at 37 °C and rinsed three times with PBS. The freshly-passivated cantilever was mounted to the AFM cantilever holder, while the material-passivated glass bottom 35 mm dishes (as described in the previous section) was installed on the AFM scanning stage. For SCFS experiments, an aliquot of $15 \times 10^3$ platelets/µl in PBS containing 1.0 mM $CaCl_2$ and 0.5 mM $MaCl_2$ was dropped onto the passivated glass right before each measurement. To immobilize a single platelet on the cantilever, the collagen passivated-cantilever was brought into contact with a loosely bound and non-activated platelet on the collagen-passivated substrate for 3–5 min. The cantilever with the adhered platelet was then moved to other immobilized platelets on each material-passivated substrate for SCFS measurement. Immediately, the collagen-passivated cantilever was brought into contact with the platelets on the passivated-substrate for immobilization of an inhibited platelet on the cantilever.

**Single-platelet force spectroscopy.**    All measurements were carried out in PBS using a JPK NanoWizard 3 (JPK, Berlin, Germany). Force distance (F-D) curves were recorded with a Z-length of 7 µm and a setpoint value of 300 pN to control the maximal force of the cantilever against the surface. A velocity of 15 µm/s was used for all measurements to avoid merging of two platelets during contact and to rupture completely two platelets from each other. For each passivated-substrate, 500 force-distance curves were taken on 7 to10 platelets. The measurements were repeated at least three times using independently freshly prepared platelets from different donors on both cantilevers and passivated-substrates. For studies with the receptor inhibitor abciximab, platelets were incubated with 50 µg/ml abciximab (Reopro, Lilly Deutschland GmbH, Giessen, Germany).

**AFM Data analysis.**    The rupture force (F) and the work area (W) were extracted directly from the retraction curve of the force-distance (F-D) curves using the JPK data processing software (version 4.4.18+). The most probable rupture force and work area values together with their corresponding errors were determined by Gaussian fits using Origin software (version 8.6, OriginLab Corporation, Northampton, MA, USA).

**Ethics.**    The use of human platelets obtained from healthy volunteers after informed consent has been approved by the University of Medicine Greifswald ethics board. The methods were carried out accordingly with the approved guidelines.

## References

1. Michelson, A. D. *Platelets*. Academic Press/Elsevier (2013).
2. Stadelmann, W. K., Digenis, A. G. & Tobin, G. R. Physiology and healing dynamics of chronic cutaneous wounds. *Am J Surg* **176**, 26S–38S (1998).
3. Nieswandt, B., Pleines, I. & Bender, M. Platelet adhesion and activation mechanisms in arterial thrombosis and ischaemic stroke. *J Thromb Haemost* **9**, 92–104 (2011).
4. Li, Z. Y., Yang, F. M. Y., Dunn, S., Gross, A. K. & Smyth, S. S. Platelets as immune mediators: Their role in host defense responses and sepsis. *Thrombosis research* **127**, 184–188 (2011).
5. Nurden, A. T., Nurden, P., Sanchez, M., Andia, I. & Anitua, E. Platelets and wound healing. *Front Biosci* **13**, 3532–3548 (2008).
6. Bhatti, R. A., Gadarowski, J. & Ray, P. Potential role of platelets and coagulation factors in the metastasis of prostatic cancer. *Invasion Metastasis* **16**, 49–55 (1996).
7. Ordinas, A., Diaz-Ricart, M., Almirall, L. & Bastida, E. The role of platelets in cancer metastasis. *Blood Coagul Fibrinolysis* **1**, 707–711 (1990).
8. Ugarova, T. P. *et al.* Conformational-Changes in Fibrinogen Elicited by Its Interaction with Platelet Membrane Glycoprotein-Gpiib-Iiia. *J Biol Chem* **268**, 21080–21087 (1993).
9. Smyth, S. S. & Parise, L. V. Regulation of Ligand-Binding to Glycoprotein Iib-Iiia (Integrin Alpha-Iib Beta-3) in Isolated Platelet Membranes. *Biochem J* **292**, 749–758 (1993).
10. Fogelson, A. L. & Neeves, K. B. Fluid Mechanics of Blood Clot Formation. *Annu Rev Fluid Mech* **47**, 377–403 (2015).
11. Whiteheart, S. W. Platelet granules: surprise packages. *Blood* **118**, 1190–1191 (2011).
12. Jackson, S. P. The growing complexity of platelet aggregation. *Blood* **109**, 5087–5095 (2007).
13. Sixma, J. J. & Wester, J. The hemostatic plug. *Semin Hematol* **14**, 265–299 (1977).
14. May, R. M. *et al.* An engineered micropattern to reduce bacterial colonization, platelet adhesion and fibrin sheath formation for improved biocompatibility of central venous catheters. *Clin Transl Med* **4**, 9 (2015).
15. Xu, L. C., Bauer, J. W. & Siedlecki, C. A. Proteins, platelets, and blood coagulation at biomaterial interfaces. *Colloid Surface B* **124**, 49–68 (2014).
16. Fritz, M., Radmacher, M. & Gaub, H. E. *In vitro* Activation of Human Platelets Triggered and Probed by Atomic Force Microscopy. *Exp Cell Res* **205**, 187–190 (1993).
17. Fritz, M., Radmacher, M. & Gaub, H. E. Granula Motion and Membrane Spreading during Activation of Human Platelets Imaged by Atomic-Force Microscopy. *Biophys J* **66**, 1328–1334 (1994).
18. Waples, L. M., Olorundare, O. E., Goodman, S. L., Lai, Q. J. & Albrecht, R. M. Platelet-polymer interactions: Morphologic and intracellular free calcium studies of individual human platelets. *J Biomed Mater Res* **32**, 65–76 (1996).
19. Nobili, M., Sheriff, J., Morbiducci, U., Redaelli, A. & Bluestein, D. Platelet activation due to hemodynamic shear stresses: Damage accumulation model and comparison to *in vitro* measurements. *Asaio J* **54**, 64–72 (2008).
20. Schafer, S. T., Neumann, A., Lindemann, J., Gorlinger, K. & Peters, J. Venous air embolism induces both platelet dysfunction and thrombocytopenia. *Acta Anaesthesiol Scand* **53**, 736–741 (2009).
21. Casa, L. D. C. & Ku, D. N. Geometric design of microfluidic chambers: platelet adhesion versus accumulation. *Biomed Microdevices* **16**, 115–126 (2014).
22. Kim, D., Finkenstaedt-Quinn, S., Hurley, K. R., Buchman, J. T. & Haynes, C. L. On-chip evaluation of platelet adhesion and aggregation upon exposure to mesoporous silica nanoparticles. *Analyst* **139**, 906–913 (2014).
23. Radomski, A. *et al.* Nanoparticle-induced platelet aggregation and vascular thrombosis. *Brit J Pharmacol* **146**, 882–893 (2005).
24. Tesfamariam, B. Platelet function in intravascular device implant-induced intimal injury. *Cardiovasc Revasc Med* **9**, 78–87 (2008).

 

25. Feghhi, S. & Sniadecki, N. J. Mechanobiology of Platelets: Techniques to Study the Role of Fluid Flow and Platelet Retraction Forces at the Micro- and Nano-Scale. *Int J Mol Sci* **12,** 9009–9030 (2011).
26. Ciciliano, J. C., Trana, R., Sakurai, Y. & Lam, W. A. The Platelet and the Biophysical Microenvironment: Lessons from Cellular Mechanics. *Thrombosis research* **133,** 532–537 (2014).
27. van Zijp, H. M. *et al.* Quantification of platelet-surface interactions in real-time using intracellular calcium signaling. *Biomed Microdevices* **16,** 217–227 (2014).
28. Lam, W. A. *et al.* Mechanics and contraction dynamics of single platelets and implications for clot stiffening. *Nat Mater* **10,** 61–66 (2011).
29. Hussain, M. A. & Siedlecki, C. A. The platelet integrin alpha(IIb) beta(3) imaged by atomic force microscopy on model surfaces. *Micron* **35,** 565–573 (2004).
30. Radmacher, M., Fritz, M., Kacher, C. M., Cleveland, J. P. & Hansma, P. K. Measuring the viscoelastic properties of human platelets with the atomic force microscope. *Biophys J* **70,** 556–567 (1996).
31. Walch, M., Ziegler, U. & Groscurth, P. Effect of streptolysin O on the microelasticity of human platelets analyzed by atomic force microscopy. *Ultramicroscopy* **82,** 259–267 (2000).
32. Friedrichs, J. *et al.* A practical guide to quantify cell adhesion using single-cell force spectroscopy. *Methods* **60,** 169–178 (2013).
33. Holland, N. B., Siedlecki, C. A. & Marchant, R. E. Intermolecular force mapping of platelet surfaces on collagen substrata. *J Biomed Mater Res* **45,** 167–174 (1999).
34. Muller, D. J. & Dufrene, Y. F. Atomic force microscopy as a multifunctional molecular toolbox in nanobiotechnology. *Nat Nanotechnol* **3,** 261–269 (2008).
35. Guo, C. L. *et al.* High-resolution probing heparan sulfate-antithrombin interaction on a single endothelial cell surface: single-molecule AFM studies. *Phys Chem Chem Phys* **17,** 13301–13306 (2015).
36. Beaussart, A. *et al.* Quantifying the forces guiding microbial cell adhesion using single-cell force spectroscopy. *Nat Protoc* **9,** 1049–1055 (2014).
37. Boettiger, D. & Wehrle-Haller, B. Integrin and glycocalyx mediated contributions to cell adhesion identified by single cell force spectroscopy. *J Phys-Condens Mat* **22** (2010).
38. Hosseini, B. H. *et al.* Immune synapse formation determines interaction forces between T cells and antigen-presenting cells measured by atomic force microscopy. *P Natl Acad Sci USA* **107,** 2373–2373 (2010).
39. Paczuski, R. Determination of von Willebrand factor activity with collagen-binding assay and diagnosis of von Willebrand disease: effect of collagen source and coating conditions. *J Lab Clin Med* **140,** 250–254 (2002).
40. Guccione, M. A., Packham, M. A., Kinlough-Rathbone, R. L., Perry, D. W. & Mustard, J. F. Reactions of polylysine with human platelets in plasma and in suspensions of washed platelets. *Thromb. Haemost* **36,** 360–375 (1976).
41. Coppinger, J. A. *et al.* Characterization of the proteins released from activated platelets leads to localization of novel platelet proteins in human atherosclerotic lesions. *Blood* **103,** 2096–2104 (2004).
42. Ma, L. *et al.* Proteinase-activated receptors 1 and 4 counter-regulate endostatin and VEGF release from human platelets. *Proc. Natl. Acad. Sci.* **102,** 216–220 (2005).
43. Nguyen, T. H., Choi, S. S., Kim, D. W. & Kim, Y. U. Study on the binding force between lambda-DNA and various types of substrates. *J Korean Phys Soc* **50,** 1942–1946 (2007).
44. Nguyen, T. H., Kim, Y. U., Kim, K. J. & Choi, S. S. Investigation of Structural Transition of dsDNA on Various Substrates Studied by Atomic Force Microscopy. *J. Nanosci. Nanotechnol.* **9,** 2162–2168 (2009).
45. Helenius, J., Heisenberg, C. P., Gaub, H. E. & Muller, D. J. Single-cell force spectroscopy. *J Cell Sci* **121,** 1785–1791 (2008).
46. Allen, R. D. *et al.* Transformation and Motility of Human-Platelets - Details of the Shape Change and Release Reaction Observed by Optical and Electron-Microscopy. *J Cell Biol* **83,** 126–142 (1979).
47. Goodman, S. L., Grasel, T. G., Cooper, S. L. & Albrecht, R. M. Platelet Shape Change and Cytoskeletal Reorganization on Polyurethaneureas. *J Biomed Mater Res* **23,** 105–123 (1989).
48. Barnhart, M. I., Walsh, R. T. & Robinson, J. A. A three-dimensional view of platelet responses to chemical stimuli. *Ann N Y Acad Sci* **201,** 360–390 (1972).
49. Nieswandt, B. & Watson, S. P. Platelet-collagen interaction: is GPVI the central receptor? *Blood* **102,** 449–461 (2003).
50. Kee, M. F., Myers, D. R., Sakurai, Y., Lam, W. A. & Qiu, Y. Platelet mechanosensing of collagen matrices. *PLos One* **10,** e0126624 (2015).
51. Jenkins, C. S., Packham, M. A., Kinlough-Rathbone, R. L. & Mustard, J. F. Interactions of polylysine with platelets. *Blood* **37,** 395–412 (1971).
52. Tiffany, M. L. & Penner, J. A. Letter: Polylysine aggregation of human blood platelets. *Thrombosis research* **8,** 529–530 (1976).
53. Ikeda, M. *et al.* Simultaneous digital imaging analysis of cytosolic calcium and morphological change in platelets activated by surface contact. *J. Cell Biochem.* **62,** U3–U4 (1996).
54. Perutelli, P., Marchese, P. & Mori, P. G. [The glycoprotein IIb/IIIa complex of the platelets. An activation-dependent integrin]. *Recenti Prog Med* **83,** 100–104 (1992).
55. Medvedev, N., Palankar, R., Krauel, K., Greinacher, A. & Delcea, M. Micropatterned array to assess the interaction of single platelets with platelet factor 4-heparin-IgG complexes. *Thromb Haemostasis* **111,** 862–872 (2014).
56. Krauel, K. *et al.* Heparin-induced thrombocytopenia - therapeutic concentrations of danaparoid, unlike fondaparinux and direct thrombin inhibitors, inhibit formation of platelet factor 4-heparin complexes. *J Thromb Haemost* **6,** 2160–2167 (2008).
57. Luthje, J. & Ogilvie, A. Platelet-Aggregation in Whole-Blood - a Laboratory Experiment for a Medical Biochemistry Course. *Biochem Educ* **15,** 86–87 (1987).
58. Zhou, L. & Schmaier, A. H. Platelet aggregation testing in platelet-rich plasma: description of procedures with the aim to develop standards in the field. *American journal of clinical pathology* **123,** 172–183 (2005).
59. Hutter, J. L. & Bechhoefer, J. Calibration of Atomic-Force Microscope Tips (Vol 64, Pg 1868, 1993). *Rev Sci Instrum* **64,** 3342–3342 (1993).
60. Lee, S. M. *et al.* Nanomechanical measurement of astrocyte stiffness correlated with cytoskeletal maturation. *J Biomed Mater Res A* **103,** 365–370 (2015).

## Acknowledgements

This work was supported by the German Ministry of Education and Research (BMBF) within the project FKZ 03Z2CN11, by the Deutsche Forschungsgemeinschaft (DFG, Germany) (NG 133/1-1), and by the European Research Council Starting Grant "PredicTOOL" (637877). We thank Ulrike Strobel and Ricarda Raschke for providing washed platelets and helping with platelet aggregation tests.

## Author Contributions

T.H.N., A.G. and M.D. conceived of the project, devised experiments and wrote the manuscript. T.H.N. and V.C.B. carried out experiments and performed data analysis for SCFS. N.M. and T.H.N. and performed SEM experiments and analyzed the data. R.P. performed and analyzed calcium imaging and immunofluorescence experiments. All the authors discussed the results, revised, and approved the final version of the manuscript.

## Additional Information

**Supplementary information** accompanies this paper at http://www.nature.com/srep

**Competing financial interests:** The authors declare no competing financial interests.

**How to cite this article**: Nguyen, T.-H. *et al.* Rupture Forces among Human Blood Platelets at different Degrees of Activation. *Sci. Rep.* **6**, 25402; doi: 10.1038/srep25402 (2016).